\documentclass[aps,pra,twocolumn,floats,floatfix]{revtex4-2}

\usepackage{amssymb}
\usepackage{amsmath}
\usepackage{calc}
\usepackage{graphicx}
\usepackage{bm}
\usepackage{comment}
\usepackage{capt-of}

\usepackage[
    colorlinks=true,
    linkcolor=blue,
    citecolor=blue,
    urlcolor=blue
]{hyperref}

\def\be{ \begin{equation} }
\def\ee{ \end{equation} }
\def\bse{ \begin{subequations} }
\def\ese{ \end{subequations} }
\def\sech{\text{sech}}

\begin{document}

\author{Aida Shiroyan}
\affiliation{Center for Quantum Technologies, Department of Physics, Sofia University, James Bourchier 5 blvd, 1164 Sofia, Bulgaria}
\author{Genko S. Vasilev}
\affiliation{Center for Quantum Technologies, Department of Physics, Sofia University, James Bourchier 5 blvd, 1164 Sofia, Bulgaria}
\author{Nikolay V. Vitanov}
\affiliation{Center for Quantum Technologies, Department of Physics, Sofia University, James Bourchier 5 blvd, 1164 Sofia, Bulgaria}

\title{Coherent excitation of a two-state system by a Lorentzian field}
\date{\today}

\begin{abstract}
We study the coherent excitation of a two-state quantum system by a field with a Lorentzian temporal envelope and constant carrier frequency. 
The associated differential equation admits an exact local representation in terms of confluent Heun functions. 
For the transition probability we develop a Dykhne-Davis-Pechukas (DDP) description based on the two relevant complex transition points and their interference. 
The DDP action is analyzed directly in the weak- and strong-coupling limits, yielding explicit asymptotic expressions for the oscillation phase, oscillation envelope, far-detuned line shape, and near-resonant behavior. 
In particular, the strong-coupling asymptotics imply a linewidth proportional to the inverse peak Rabi frequency for resonant odd-$\pi$ pulses, thereby exhibiting the power-narrowing characteristic of the Lorentzian pulse.
\end{abstract}

\maketitle

\section{Introduction}

Driven two-state systems are among the fundamental models of quantum mechanics and occur in nuclear magnetic resonance, coherent atomic and molecular excitation, quantum information, condensed-matter physics, chemical physics, and neutrino oscillations. Moreover, many multistate problems with more complicated coupling structures can be reduced, exactly or approximately, to one or more effective two-state subsystems \cite{Shore90}. Although a general closed-form solution for an arbitrary time-dependent two-state Hamiltonian is not available, a number of exactly soluble models play a central role in quantum dynamics, including the Rabi \cite{Rabi}, Landau-Majorana-St\"uckelberg-Zener \cite{Landau1932,Majorana1932,Stueckelberg1932,Zener1932}, Rosen-Zener \cite{RZ}, Allen-Eberly-Hioe \cite{AllenEberly1975,HioeEberly1981}, Bambini-Berman \cite{BB}, Demkov-Kunike \cite{DK}, Demkov \cite{Demkov}, and Nikitin \cite{Nikitin} models. Further extensions include exact treatments of arbitrary amplitude and frequency modulations \cite{HioeCarroll1985} and finite-duration corrections to the Landau-Zener problem \cite{VitanovGarraway1996}.

The Lorentzian pulse is particularly interesting because this shape is widely used in physics in various contexts.
The corresponding differential equation, stemming from the Schr\"odinger equation, can be mapped onto a confluent Heun equation. 
Confluent-Heun solutions of broad classes of two-state models, including pulse families closely related to the Lorentzian model, have been classified in a more general framework \cite{IshkhanyanGrigoryan2014}; exact analytical results for a two-level system driven by a Lorentzian-shaped pulse have also been revisited in later work \cite{XieLiu2020}. 
We therefore do not base the novelty of the present analysis on the mere existence of a Heun representation. 
Our emphasis is instead on the complex-time structure of the Lorentzian model, the interference of its two relevant transition points, and the physical information that can be extracted analytically from the DDP approach.

The strong pulse-shape dependence of post-pulse two-state excitation was analyzed previously for Lorentzian, squared-Lorentzian, Gaussian, hyperbolic-secant, and squared-hyperbolic-secant envelopes \cite{Berman1998}, and was subsequently demonstrated experimentally for the same family of pulse shapes, including the Lorentzian and squared-Lorentzian cases \cite{Conover2011}. 
Closely related DDP analyses of a Gaussian envelope with constant carrier frequency \cite{VasilevVitanov2004} and of a linearly chirped Gaussian pulse \cite{VasilevVitanov2005} provide useful comparisons with the meromorphic Lorentzian model considered here. 
Finite-duration effects for shaped pulses have likewise been analyzed systematically \cite{BoradjievVitanov2013Finite,MihovVitanov2024Finite}, while approximate analytic treatments of sinusoidal near-resonant dynamics provide another comparison with the Lorentzian case \cite{YatsenkoGuerinJauslin2004}.

The algebraic pulse wings also make the Lorentzian model important for spectral selectivity. For pulses with tails that vanish as an inverse power of time, adiabatic arguments predict power narrowing rather than conventional power broadening: the post-pulse excitation linewidth decreases as the peak Rabi frequency increases \cite{BoradjievVitanov2013}. This effect was later demonstrated experimentally on a superconducting quantum processor for powers of Lorentzian pulses, including the ordinary Lorentzian shape \cite{MihovVitanov2024}. Conversely, other smooth envelopes can exhibit power superbroadening, with the linewidth increasing strongly at successive excitation maxima \cite{MihovVitanov2025Superbroadening}. One aim of the asymptotic analysis below is to show directly how the same inverse-Rabi-frequency linewidth scaling emerges from the DDP action of the Lorentzian model.
We also compare the Lorentzian result with the Rosen-Zener conjecture and its Robiscoe-type detuning-dependent-area generalization, which have proved accurate for several other smooth pulse shapes \cite{Robiscoe1983,Thomas83,MihovVitanov2023}.

The broader physical context is the suppression of conventional power broadening when a coherently excited population is measured after a smooth pulse rather than continuously during the interaction \cite{VitanovEtAl2001,HalfmannEtAl2003}. The present calculation concerns an ideal Lorentzian pulse with infinite algebraic wings; more general shaped fields may also drive substantial higher-order transitions even when their spectra contain no component at the Bohr frequency \cite{VitanovShore2005}. Different amplitude-detuning pairs can also be related through isoprobability transformations \cite{HioeCarroll1985JOSAB,MihovVitanov2025Isoprobability}; here, however, the detuning is held constant and the analysis focuses on the Lorentzian envelope itself.

The paper is organized as follows. 
Section~\ref{Sec-background} defines the model and the dimensionless parameters. 
Section~\ref{Sec-Heun} discusses the reduction to a confluent Heun equation and clarifies the status of the local exact solution. 
Section~\ref{Sec-DDP-main} develops the DDP description, including the transition points, complex singularities, action integral, and transition probability. 
Section~\ref{Sec-asymptotics} derives the weak- and strong-coupling asymptotics, compares the Lorentzian result with the Rosen-Zener-Robiscoe conjecture, and discusses their implications for the line shape, oscillation period, and power narrowing.
Section~\ref{Sec-Pade} develops the minimal $[3/1]$ and constrained $[4/2]$ two-point Pad\'e approximations.
The main results are summarized in Sec.~\ref{Sec-conclusion}.

\section{Basic equations and definitions\label{Sec-background}}

The probability amplitudes in the diabatic basis, $\mathbf{c}(t)=[c_1(t),c_2(t)]^T$, satisfy
\begin{equation}
 i\hbar\frac{d}{dt}\mathbf{c}(t)=\mathbf{H}(t)\mathbf{c}(t),
 \label{Schrodinger-2SS}
\end{equation}
with the rotating-wave Hamiltonian
\begin{equation}
\mathbf{H}(t)=\frac{\hbar}{2}
\begin{pmatrix}
-\Delta & \Omega(t)\\
\Omega(t) & \Delta
\end{pmatrix}.
\label{H2}
\end{equation}
Here $\Delta=\omega_0-\omega$ is the detuning of the carrier frequency $\omega$ from the Bohr transition frequency $\omega_0$, and $\Omega(t)$ is the Rabi frequency. We consider
\begin{subequations}
\label{model}
\begin{align}
\Omega(t)&=\frac{\Omega_0}{1+t^2/T^2},\label{Rabi}\\
\Delta(t)&=\Delta=\mathrm{const}.\label{detuning}
\end{align}
\end{subequations}
Because the transition probability is even in $\Omega_0$ and $\Delta$, we take $\Omega_0>0$, $\Delta>0$, and $T>0$ unless stated otherwise.

For an initial state $|\psi_1\rangle$, we have
$c_1(-\infty)=1, c_2(-\infty)=0$ and the post-pulse transition probability is
$\mathcal{P}=|c_2(+\infty)|^2$.
It is convenient to introduce the dimensionless variables
\begin{equation}
\tau=\frac{t}{T},\qquad \alpha=\Omega_0T,
\qquad \delta=\Delta T,
\label{dimensionless}
\end{equation}
and the ratio
\begin{equation}
\beta=\frac{\Omega_0}{\Delta}=\frac{\alpha}{\delta}.
\label{beta-ratio}
\end{equation}
The Lorentzian pulse area is
$A=\int_{-\infty}^{\infty}\Omega(t)\,dt=\pi\Omega_0T=\pi\alpha$;
hence $\alpha=A/\pi$ measures the pulse area in units of $\pi$.
The dynamics therefore depends only on two independent dimensionless parameters, for example $(\alpha,\delta)$, while $\beta=\alpha/\delta$ controls the coupling-to-detuning ratio.

\section{Reduction to the confluent Heun equation\label{Sec-Heun}}

For the reduction to a scalar equation it is convenient to remove an irrelevant common energy shift. Multiplying the state vector by a global phase is equivalent to replacing Eq.~(\ref{H2}) by
\begin{equation}
\mathbf{H}'(t)=\mathbf{H}(t)+\frac{\hbar\Delta}{2}\mathbf{1}
=\hbar
\begin{pmatrix}
0 & \Omega(t)/2\\
\Omega(t)/2 & \Delta
\end{pmatrix},
\end{equation}
which leaves all transition probabilities unchanged. Eliminating the second amplitude gives
\begin{equation}
\frac{d^2c_1}{dt^2}
+\left(i\Delta-\frac{\dot\Omega}{\Omega}\right)\frac{dc_1}{dt}
+\frac{\Omega^2}{4}c_1=0.
\label{scalar-dimensional}
\end{equation}
For the Lorentzian pulse and the dimensionless variables of Eq.~(\ref{dimensionless}), this becomes
\begin{equation}
\frac{d^2y}{d\tau^2}
+\left(\frac{2\tau}{1+\tau^2}+i\delta\right)\frac{dy}{d\tau}
+\frac{\alpha^2}{4(1+\tau^2)^2}y=0,
\label{DELrtz}
\end{equation}
with $y(\tau)=c_1(t)$.
Thus all parameters appearing in the scalar differential equation are dimensionless.

Equation~(\ref{DELrtz}) can be mapped to a confluent Heun equation, or equivalently to a generalized spheroidal wave equation. A convenient form of the latter is
\begin{equation}
\begin{aligned}
 z(z-z_0)\frac{d^2u}{dz^2}
 &+\left(B_1+B_2z\right)\frac{du}{dz}\\
 &+\left[B_3-2\eta\omega(z-z_0)
 +\omega^2z(z-z_0)\right]u=0,
\end{aligned}
\label{GSWE}
\end{equation}
where the two finite singular points are regular and infinity is irregular \cite{Leaver}. The latter feature is important here because the physical boundary conditions are imposed at $\tau=\pm\infty$.

To make the special-function convention explicit, define
\begin{equation}
 z=\frac{1-i\tau}{2}
\label{Heun-z}
\end{equation}
and let $\mathrm{CHl}(\sigma;\kappa,\gamma_{\rm H},\eta_{\rm H},\epsilon_{\rm H};z)$ denote the Frobenius solution about $z=0$, normalized to unity there, of
\begin{equation}
\frac{d^2u}{dz^2}
+\left(\frac{\gamma_{\rm H}}{z}
+\frac{\eta_{\rm H}}{z-1}+\epsilon_{\rm H}\right)\frac{du}{dz}
+\frac{\kappa z-\sigma}{z(z-1)}u=0.
\label{canonical-CHE}
\end{equation}
The subscripts ``H'' distinguish the canonical Heun parameters from the physical detuning and the dimensionless variables used elsewhere.

Writing
\begin{equation}
 y(\tau)=e^{-i\delta\tau}
 \left(\frac{\tau-i}{\tau+i}\right)^{\alpha/4}u(z),
\label{Heun-substitution}
\end{equation}
one obtains the local basis
\begin{equation}
\begin{aligned}
 y(\tau)={}&e^{-i\delta\tau}
 \left(\frac{\tau-i}{\tau+i}\right)^{\alpha/4}\\
 &\times\left[C_1(\tau+i)^{\alpha/2}\mathrm{CHl}_1(z)
 +C_2\mathrm{CHl}_2(z)\right],
\end{aligned}
\label{Heun-local-solution}
\end{equation}
where constant numerical factors from the change of variable have been absorbed into $C_1$ and $C_2$. The two local Heun functions are
\begin{subequations}
\label{CHl-definitions}
\begin{align}
\mathrm{CHl}_1(z)&=\mathrm{CHl}
 (\sigma_1;\kappa_1,\gamma_1,\eta_1,\epsilon_1;z),\\
\mathrm{CHl}_2(z)&=\mathrm{CHl}
 (\sigma_2;\kappa_2,\gamma_2,\eta_2,\epsilon_2;z),
\end{align}
\end{subequations}
with
\begin{subequations}
\label{CHl-parameters}
\begin{align}
\sigma_1&=\frac{-\alpha^2+2\alpha\delta-2\alpha+8\delta}{4},
&\kappa_1&=\delta(\alpha+4),\\
\gamma_1&=1+\frac{\alpha}{2},
&\eta_1&=1+\frac{\alpha}{2},
&\epsilon_1&=2\delta,\\
\sigma_2&=\frac{\delta(4-\alpha)}{2},
&\kappa_2&=4\delta,\\
\gamma_2&=1-\frac{\alpha}{2},
&\eta_2&=1+\frac{\alpha}{2},
&\epsilon_2&=2\delta.
\end{align}
\end{subequations}

Although Eq.~(\ref{Heun-local-solution}) provides an exact local representation of the solution, it does not directly determine the physical scattering amplitude. 
The limits $\tau\rightarrow-\infty$ and $\tau\rightarrow+\infty$ correspond to approaching the irregular
singularity of the confluent Heun equation from different asymptotic
sectors. Determining the transition probability from the Heun
representation would therefore require a connection matrix relating
the solution selected by the initial conditions
$c_1(-\infty)=1$ and $c_2(-\infty)=0$ to the asymptotic basis at
$+\infty$, together with the appropriate Stokes continuation. No
simple closed-form connection formula is available for this geometry.
A numerical continuation of the Heun functions would in practice offer
no advantage over direct numerical propagation of the original
Schr\"odinger equation. We therefore use the Heun representation to
establish the exact special-function class of the model, while the
transition probability and its physically transparent asymptotic
behavior are obtained below by the DDP method.

\section{Dykhne-Davis-Pechukas analysis\label{Sec-DDP-main}}

\subsection{Adiabatic basis}

Define the mixing angle by
\begin{equation}
\tan 2\vartheta(\tau)=\frac{\Omega(\tau)}{\Delta},
\qquad 0\leq\vartheta(\tau)\leq\frac{\pi}{4}.
\label{theta}
\end{equation}
The instantaneous adiabatic states are
\begin{subequations}
\begin{align}
|\varphi_-(\tau)\rangle&=\cos\vartheta(\tau)|\psi_1\rangle-\sin\vartheta(\tau)|\psi_2\rangle,\\
|\varphi_+(\tau)\rangle&=\sin\vartheta(\tau)|\psi_1\rangle+\cos\vartheta(\tau)|\psi_2\rangle.
\end{align}
\end{subequations}
Since $\Omega(\tau)\rightarrow0$ for $\tau\rightarrow\pm\infty$ and $\Delta>0$, one has $\vartheta(\pm\infty)=0$, so the adiabatic and diabatic bases coincide asymptotically.

The eigenvalues of the Hamiltonian in Eq.~(\ref{H2}) are
\begin{equation}
E_\pm(\tau)=\pm\frac{\hbar}{2}\sqrt{\Omega^2(\tau)+\Delta^2},
\label{adiabatic-eigenvalues}
\end{equation}
and the quasienergy splitting is
\begin{equation}
\hbar\mathcal{E}(\tau)=E_+(\tau)-E_-(\tau)
=\hbar\sqrt{\Omega^2(\tau)+\Delta^2}.
\label{splitting}
\end{equation}
In the adiabatic basis the nonadiabatic coupling is proportional to $d\vartheta/d\tau$.

\subsection{Single and multiple transition points}

For a single transition point $t_0$, defined as $\mathcal{E}(t_0)=0$, situated in the upper half of the complex-time plane, the DDP approximation gives \cite{Dykhne,Davis76}
\begin{equation}
\mathcal{P}\sim e^{-2\,\mathrm{Im}\,\mathcal{D}(t_0)},
\end{equation}
where, writing $t_0=T\tau_0$,
\begin{equation}
\mathcal{D}(t_0)=\int_0^{t_0}\mathcal{E}(t)\,dt
=T\int_0^{\tau_0}\mathcal{E}(\tau)\,d\tau.
\label{D}
\end{equation}
Because $\hbar\mathcal{E}(t)=E_+(t)-E_-(t)$, the dimensional quantity
\begin{equation}
\hbar\mathcal{D}(t_0)=\int_0^{t_0}\left[E_+(t)-E_-(t)\right]dt
\label{dimensional-DDP-action}
\end{equation}
has the dimensions of action. We therefore refer to $\mathcal{D}$ as the dimensionless complex DDP action, or equivalently as the analytically continued relative adiabatic phase integral. This terminology reflects its semiclassical role: the transition amplitudes contain $e^{i\mathcal{D}}=e^{-\operatorname{Im}\mathcal{D}}e^{i\operatorname{Re}\mathcal{D}}$, so $\operatorname{Re}\mathcal{D}$ controls the interference phase, whereas $\operatorname{Im}\mathcal{D}$ controls the exponential attenuation. The DDP action should not be confused with the full classical mechanical action; it is specifically the integral of the adiabatic energy splitting along a generally complex-time contour.

When several transition points $\tau_k=t_k/T$ contribute on the relevant Stokes manifold, their amplitudes must be summed coherently,
\begin{equation}
\mathcal{P}\sim\left|\sum_k\Gamma(\tau_k)e^{i\mathcal{D}(\tau_k)}\right|^2,
\label{DP-N}
\end{equation}
with the dimensionless prefactors
\begin{equation}
\Gamma(\tau_k)=4i\lim_{\tau\rightarrow\tau_k}
(\tau-\tau_k)\frac{d\vartheta}{d\tau}.
\label{Gamma-k}
\end{equation}
The transition-point prefactors and their universality for different local singularity structures have been analyzed systematically using superadiabatic renormalization \cite{BerryLim1993}. The need to retain a pair of complex singularities, rather than only the nearest one, already appears in the adiabatic analysis of level-crossing models \cite{SuominenGarrawayStenholm1991}; the general interference formula for several complex crossings is given in Ref.~\cite{Joye91}.
The usual DDP assumptions include analyticity in the relevant complex domain, an appropriate choice of branches and contours, and the existence of Stokes lines connecting the real-time evolution to the contributing transition points \cite{Davis76,Joye91}.

\subsection{Transition points, poles, and Stokes structure\label{Sec-transition-points}}

For the Lorentzian model the upper-half-plane transition points are
\begin{equation}
\tau^\pm=
\pm\sqrt{\frac{-1+\sqrt{1+\beta^2}}{2}}
+i\sqrt{\frac{1+\sqrt{1+\beta^2}}{2}}.
\label{trans-points}
\end{equation}
They satisfy the symmetry relations
\begin{subequations}
\label{transition-point-relations}
\begin{align}
1+(\tau^\pm)^2&=\pm i\beta,\\
(\tau^-)^*&=-\tau^+.
\end{align}
\end{subequations}
The Lorentzian coupling itself has simple poles at
\begin{equation}
\tau_0^\pm=\pm i.
\label{Lorentzian-poles}
\end{equation}
For every $\beta>0$, $\mathrm{Im}\,\tau^+>1$, so the pole at $\tau_0^+=i$ lies closer to the real axis than the transition points. This feature distinguishes the Lorentzian model from entire pulse shapes such as the Gaussian and must be accounted for when defining the analytic continuation of the quasienergy and the integration contour. Modern complex-contour analyses of the DDP method emphasize that poles obstructing a contour deformation must be treated explicitly rather than passed through implicitly \cite{FukushimaShimazaki2020}. The physical branch of $\mathcal{E}(\tau)$ is chosen to be positive on the real axis. We define the contour $\mathcal{C}_+$ by analytic continuation from the origin into the first quadrant to $\tau^+$, within the homotopy class that passes to the right of the pole at $\tau=i$ and crosses neither a pole nor a branch cut. The contour to the second-quadrant transition point is fixed by symmetry, $\mathcal{C}_-=-\mathcal{C}_+^*$. Branch cuts are taken to emanate from the transition points away from these contours and away from the real axis. 
With this prescription, the two actions are evaluated on mutually consistent sheets.

\begin{figure}[tbph]
\includegraphics[width=0.95\columnwidth]{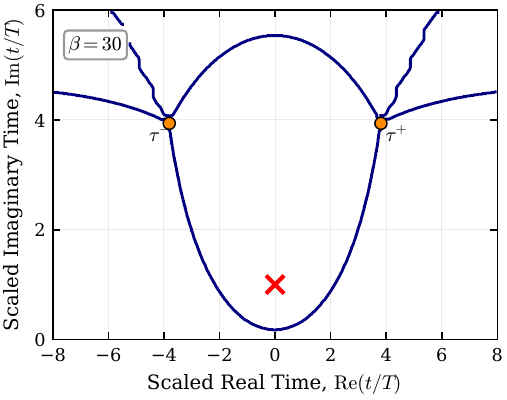}
\caption{Representative Stokes-line structure for the Lorentzian model at $\beta=30$ in the upper half of the scaled complex-time plane $t/T$. The plotted curves are the numerical Stokes trajectories issuing from $\tau^\pm$, obtained from the level-set condition $\operatorname{Im}\int_{\tau^\pm}^{\tau}\mathcal{E}(\tau')\,d\tau'=\mathrm{const}$ on the physical sheets. The two transition points lie on the dominant Stokes manifold relevant to the generalized DDP interference. The Lorentzian pole at $t/T=i$ is marked separately and must not be crossed by the analytic-continuation contours.}
\label{stokes}
\end{figure}

Because the quasienergy splitting is even in time, we have
\begin{equation}
\mathcal{D}(\tau^-)=-\mathcal{D}^*(\tau^+),
\label{D-D}
\end{equation}
so the real parts of the two actions have opposite signs and the imaginary parts are equal. The Stokes geometry confirms that the two transition points contribute coherently to the same physical transition amplitude.

\subsection{DDP action integral}

Using $\tau=t/T$, the integral for the upper-right transition point is
\begin{equation}
\mathcal{D}(\tau^+)=
\Delta T\int_0^{\tau^+}
\frac{\sqrt{\beta^2+(1+\tau^2)^2}}{1+\tau^2}\,d\tau.
\label{D(Tc)}
\end{equation}
The square root is defined by continuity from the positive quasienergy on the real axis. Equation~(\ref{D(Tc)}) is the primary definition used throughout this work.

The integral belongs to the elliptic class and can be represented in terms of incomplete elliptic integrals of the first, second, and third kinds \cite{EllipticIntegrals}. With the parameter convention $m$ these are
\begin{subequations}
\begin{align}
F(\phi,m)&=\int_0^\phi\frac{d\theta}{\sqrt{1-m\sin^2\theta}},\\
E(\phi,m)&=\int_0^\phi\sqrt{1-m\sin^2\theta}\,d\theta,\\
\Pi(n;\phi,m)&=\int_0^\phi
\frac{d\theta}{(1-n\sin^2\theta)\sqrt{1-m\sin^2\theta}}.
\end{align}
\label{elliptic-definitions}
\end{subequations}
For complex arguments, however, closed forms in terms of $F$, $E$, and $\Pi$ depend explicitly on branch conventions and analytic-continuation paths. To avoid obscuring the physical branch by a convention-dependent formula, we use the contour integral (\ref{D(Tc)}) directly for the asymptotic analysis and for numerical evaluation. This representation also has the advantage that its limiting behavior is unambiguous.

\subsection{Transition probability}

For the two transition points of Eq.~(\ref{trans-points}), one finds
\begin{equation}
\Gamma(\tau^\pm)=\pm1.
\label{Gamma-k-1}
\end{equation}
Using Eq.~(\ref{D-D}), the generalized DDP expression and its unitarized interpolation are
\begin{subequations}
\label{P-DDP-pair}
\begin{align}
\mathcal{P}_{\mathrm{DDP}}^{(\mathrm{s})}
&\sim4e^{-2\,\mathrm{Im}\mathcal{D}(\tau^+)}
\sin^2\!\left[\mathrm{Re}\mathcal{D}(\tau^+)\right],
\label{P-DDP}\\
\mathcal{P}_{\mathrm{DDP}}^{(\mathrm{u})}
&\sim
\frac{\sin^2[\mathrm{Re}\mathcal{D}(\tau^+)]}
{\cosh^2[\mathrm{Im}\mathcal{D}(\tau^+)]}.
\label{P-DDP-sech}
\end{align}
\end{subequations}
The standard expression is asymptotically justified in the DDP regime but can exceed unity when extrapolated outside it.
For $\mathrm{Im}\mathcal{D}\gg1$, $\sech^2x\simeq4e^{-2x}$ and Eq.~(\ref{P-DDP-sech}) reduces to Eq.~(\ref{P-DDP}). The replacement is therefore asymptotically consistent while enforcing $0\leq\mathcal{P}\leq1$; related unitarizing substitutions have been used previously to extend semiclassical transition formulas beyond their strict asymptotic domain \cite{Crothers}. We will distinguish below between conclusions that follow from the DDP action itself and those that additionally rely on the unitarized interpolation.

\section{Asymptotic analysis of the DDP action\label{Sec-asymptotics}}

By the exact symmetry relation of Eq.~(\ref{D-D}), $\mathcal{D}(\tau^-)$ carries the same information as $\mathcal{D}(\tau^+)$ but admits an algebraically simpler closed form in terms of complete elliptic integrals, derived in Appendix~\ref{App-elliptic}:
\begin{subequations}
\label{eq:Dtauminus-group}
\begin{align}
\mathcal{D}(\tau^-)={}&\lambda\big[(i+\beta)E(m)-\beta K(m)
+i\beta^2\Pi(1{+}i\beta,m)\big],
\label{eq:Dtauminus}\\
\lambda&\equiv\frac{\Delta T}{\sqrt{1-i\beta}},
\qquad
m\equiv\frac{i-\beta}{i+\beta}.
\label{eq:DtauminusCoe}
\end{align}
\end{subequations}
The weak- and strong-coupling expansions of this complete-elliptic representation are presented together with the corresponding $\tau^+$ expansions in the relevant subsections below. The representation is used only on the branch fixed by the contour prescription above and does not define an independent, inequivalent action.

The direct integral representation in Eq.~(\ref{D(Tc)}) makes it possible to obtain compact asymptotic formulas without separately expanding branch-dependent elliptic functions. The ratio $\beta=\alpha/\delta$ controls the transition-point geometry. The limit $\beta\ll1$ corresponds either to weak coupling at fixed detuning or to large detuning at fixed coupling, whereas $\beta\gg1$ describes either strong coupling or small detuning.

\subsection{Weak coupling and large detuning\label{sec:weak}}

For $\beta\ll1$, the upper transition point approaches the Lorentzian pole at $\tau_0^+=i$,
\begin{equation}
\tau^+=i+\frac{\beta}{2}+\frac{i\beta^2}{8}
-\frac{\beta^3}{16}+O(\beta^4).
\label{tau-small-beta}
\end{equation}
The DDP actions on the two physical branches have the paired weak-coupling expansions
\begin{subequations}
\label{D-small-beta-pair}
\begin{align}
\frac{\mathcal{D}(\tau^+)}{\Delta T}\simeq{}&
 i+\frac{\pi}{4}\beta+\frac{\pi}{16}\beta^2
 +i\frac{\beta^2}{8}
\left[\ln\left(\frac{8}{\beta}\right)-\frac12\right]+\cdots,
\label{D-small-beta}\\
\frac{\mathcal{D}(\tau^-)}{\Delta T}\simeq{}&
 i-\frac{\pi}{4}\beta-\frac{\pi}{16}\beta^2
+i\frac{\beta^2}{8}
\left[\ln\left(\frac{8}{\beta}\right)-\frac12\right]+\cdots.
\label{eq:Dminus-weak}
\end{align}
\end{subequations}
The second expression is $-\mathcal{D}^*(\tau^+)$, as required by Eq.~(\ref{D-D}).
Hence
\begin{subequations}
\begin{align}
\mathrm{Re}\,\mathcal{D}
&\simeq\frac{\pi}{4}\Omega_0T
+\frac{\pi}{16}\frac{\Omega_0^2T}{\Delta},\label{D-small-real}\\
\mathrm{Im}\,\mathcal{D}
&\simeq\Delta T
+\frac{\Omega_0^2T}{8\Delta}
\left[\ln\left(\frac{8\Delta}{\Omega_0}\right)-\frac12\right].
\label{D-small-imag}
\end{align}
\end{subequations}
The dominant imaginary part is $\Delta T$, which produces exponentially suppressed far-detuned wings. To leading order,
\begin{equation}
\mathcal{P}_{\mathrm{DDP}}
\sim4e^{-2\Delta T}
\sin^2\left(\frac{\pi\Omega_0T}{4}\right),
\quad (\Omega_0/\Delta\ll1).
\label{P-large-detuning-leading}
\end{equation}
Keeping the first corrections gives
\begin{equation}
\begin{aligned}
\mathcal{P}_{\mathrm{DDP}}\simeq{}&4\exp\Bigg\{-2\Delta T
-\frac{\Omega_0^2T}{4\Delta}\left[\ln\left(\frac{8\Delta}{\Omega_0}\right)-\frac12\right]\Bigg\}\\
&\times\sin^2\left[
\frac{\pi\Omega_0T}{4}
+\frac{\pi\Omega_0^2T}{16\Delta}\right].
\end{aligned}
\label{P-large-detuning-corrected}
\end{equation}

In the additional weak-pulse limit $\alpha=\Omega_0T\ll1$,
\begin{equation}
\mathcal{P}\simeq
\frac{\pi^2}{4}(\Omega_0T)^2e^{-2\Delta T}.
\label{P-weak}
\end{equation}
This agrees with first-order perturbation theory because (for the assumed $\Delta>0$)
\begin{equation}
\frac{\Omega_0}2
\int_{-\infty}^{\infty}
\frac{e^{i\Delta t}}{1+t^2/T^2}\,dt
=\frac{\pi \Omega_0 T}{2} e^{-\Delta T}.
\label{Lorentzian-FT}
\end{equation}
Thus the DDP asymptotics reproduce the exact weak-field Fourier-tail dependence of the Lorentzian pulse. This Fourier-transform interpretation applies to the leading perturbative regime; beyond it, coherent higher-order processes can produce appreciable excitation even for fields lacking a resonant spectral component \cite{VitanovShore2005}.

\subsection{Strong coupling and small detuning\label{sec:strong}}

For $\beta\gg1$,
\begin{equation}
\tau^\pm\simeq(\pm1+i)\sqrt{\frac{\beta}{2}},
\label{tau-large-beta}
\end{equation}
so the transition points recede to infinity as $\Delta\rightarrow0$. 
The corresponding two branch expansions are
\begin{subequations}
\label{D-large-beta-pair}
\begin{align}
\frac{\mathcal{D}(\tau^+)}{\Delta T}
&\simeq\frac{\pi}{2}\beta+(-1+i)C\sqrt{\beta}
+O(\beta^{-1/2}),
\label{D-large-beta}\\
\frac{\mathcal{D}(\tau^-)}{\Delta T}
&\simeq-\frac{\pi}{2}\beta+(1+i)C\sqrt{\beta}
+O(\beta^{-1/2}),
\label{eq:Dminus-strong}\\
C&=\frac{\Gamma^2(3/4)}{\sqrt{\pi}}
=0.8472130848\ldots.
\label{C-constant}
\end{align}
\end{subequations}
The two expansions satisfy Eq.~(\ref{D-D}) term by term.
Equivalently,
\begin{subequations}
\begin{align}
\mathrm{Re}\,\mathcal{D}
&\simeq\frac{\pi}{2}\Omega_0T
-CT\sqrt{\Omega_0\Delta},\label{D-large-real}\\
\mathrm{Im}\,\mathcal{D}
&\simeq CT\sqrt{\Omega_0\Delta}.
\label{D-large-imag}
\end{align}
\end{subequations}
The real part gives the interference phase, while the imaginary part determines the attenuation of the two transition-point contributions. The leading phase $\pi\Omega_0T/2$ is one half of the Lorentzian pulse area
\begin{equation}
A=\int_{-\infty}^{\infty}\Omega(t)\,dt=\pi\Omega_0T.
\end{equation}
The first detuning correction scales as $\sqrt{\Omega_0\Delta}$.

The standard DDP expression and the unitarized near-resonant form are
\begin{subequations}
\label{P-strong-pair}
\begin{align}
\mathcal{P}_{\mathrm{DDP}}^{(\mathrm{s})}
\sim{}&4\exp[-2CT\sqrt{\Omega_0\Delta}]
\notag\\
&\times\sin^2\left[
\frac{\pi}{2}\Omega_0T-CT\sqrt{\Omega_0\Delta}\right],
\label{P-large-DDP}\\
\mathcal{P}_{\mathrm{DDP}}^{(\mathrm{u})}
\sim{}&\sech^2\left(CT\sqrt{\Omega_0|\Delta|}\right)
\notag\\
&\times\sin^2\left[
\frac{\pi}{2}\Omega_0T-CT\sqrt{\Omega_0|\Delta|}\right].
\label{P-small-detuning}
\end{align}
\end{subequations}
The ordinary DDP prefactor is not uniform at exact resonance because $\mathrm{Im}\,\mathcal{D}\rightarrow0$; the unitarized expression is therefore more useful for describing the approach to resonance.
The absolute value emphasizes the symmetry of the transition probability under $\Delta\rightarrow-\Delta$. Equation~(\ref{P-small-detuning}) should be regarded as a near-resonant asymptotic interpolation: the square-root structure follows from the DDP action, whereas the precise probability prefactor relies on the unitarized continuation in Eq.~(\ref{P-DDP-sech}).


In terms of $\alpha=\Omega_0T$ and $\delta=\Delta T$, the strong-coupling phase and the associated local oscillation period are
\begin{subequations}
\label{phase-period-group}
\begin{align}
\Phi(\alpha,\delta)
&\simeq\frac{\pi}{2}\alpha-C\sqrt{\alpha\delta},
\label{phase-alpha-delta}\\
\Delta\alpha_{\mathrm{osc}}
&\simeq\frac{\pi}{d\Phi/d\alpha}
=\frac{2}{1-(C/\pi)\sqrt{\delta/\alpha}},
\label{period-large}\\
\Delta\alpha_{\mathrm{osc}}
&\simeq2\left[1+\frac{C}{\pi}\sqrt{\frac{\delta}{\alpha}}
+O\left(\frac{\delta}{\alpha}\right)\right],
\quad \alpha\gg\delta.
\label{period-large-expanded}
\end{align}
\end{subequations}
The period therefore approaches the exact resonant value $\Delta \alpha=2$ from above.

\subsection{Power narrowing from the strong-coupling asymptotics}
\label{Sec-power-narrowing}

The strong-coupling result also reveals directly the power-narrowing behavior of the Lorentzian pulse. For the resonant maxima and the corresponding scaling variable,
\begin{subequations}
\label{power-narrowing-core}
\begin{align}
\alpha&=\Omega_0T=2N+1,
\qquad N=0,1,2,\ldots,
\label{odd-pi-pulses}\\
x&=C\sqrt{\alpha|\delta|}
=CT\sqrt{\Omega_0|\Delta|},\\
\mathcal{P}_{\mathrm{DDP}}
&\sim\sech^2x\,\cos^2x.
\label{P-power-narrowing-scaling}
\end{align}
\end{subequations}
The exact resonant probability is unity for every member of the odd-$\pi$ sequence.
The important point is that the detuning and Rabi frequency enter this profile only through the product $\alpha|\delta|$. Therefore any fixed probability level in the central line profile corresponds asymptotically to
\begin{equation}
\alpha|\delta|=\mathrm{const}.
\end{equation}
Let $x_{1/2}$ be the smallest positive solution of $\sech^2x_{1/2}\cos^2x_{1/2}=1/2$. Numerically, $x_{1/2}=0.5871381494\ldots$, and Eq.~(\ref{P-power-narrowing-scaling}) therefore predicts
\begin{equation}
|\delta_{1/2}|\simeq\frac{x_{1/2}^2}{C^2\alpha}
=\frac{0.4802808716\ldots}{\alpha}.
\label{power-narrowing-prefactor}
\end{equation}
Equivalently, if $\Delta_{1/2}$ denotes the positive half-width at half maximum,
\begin{subequations}
\label{power-narrowing-scaling}
\begin{align}
\Delta_{1/2}T&\propto\frac{1}{\Omega_0T},\\
\Delta_{1/2}&\propto\frac{1}{\Omega_0T^2}.
\end{align}
\end{subequations}
Thus, at fixed pulse duration, the linewidth decreases inversely with the peak Rabi frequency. This is precisely the power-narrowing scaling previously predicted from adiabatic population-return arguments for a Lorentzian tail, $\Omega(t)\sim|t|^{-2}$ \cite{BoradjievVitanov2013}. More generally, for tails $|t|^{-\lambda}$ the predicted scaling is $\Delta_{1/2}T\propto(\Omega_0T)^{-1/(\lambda-1)}$; setting $\lambda=2$ gives the $1/\Omega_0$ law recovered here. The same qualitative effect, including pronounced narrowing of Lorentzian and powers-of-Lorentzian excitation profiles with increasing pulse area, has since been observed experimentally on IBM Quantum hardware \cite{MihovVitanov2024}.

Equation~(\ref{power-narrowing-scaling}) is significant because it emerges here from the complex-time DDP action rather than directly from the real-time adiabaticity criterion. The two approaches therefore provide complementary descriptions of the same physical mechanism: increasing the peak coupling pushes the nonadiabatic region into a progressively narrower interval of detunings, while the algebraic pulse wings maintain adiabatic population return farther from resonance. The distinction between excitation observed during and after a smooth coherent pulse, and the associated elimination or suppression of ordinary power broadening, was established theoretically and experimentally in Refs.~\cite{VitanovEtAl2001,HalfmannEtAl2003}. We stress, however, that the scaling law is more robust than the numerical prefactor inferred from Eq.~(\ref{P-power-narrowing-scaling}), because the latter uses the unitarized DDP interpolation close to resonance.

\begin{figure}[tbph]
\centering
\includegraphics[width=1\columnwidth]{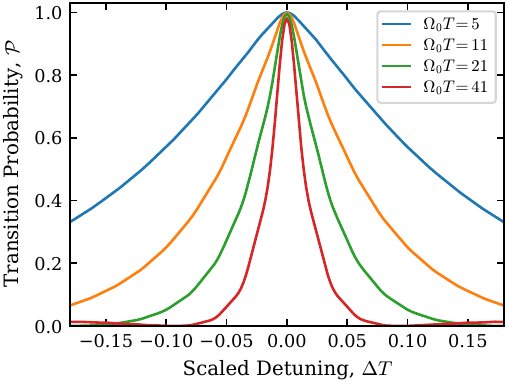}
\caption{Numerically integrated transition probability versus the scaled detuning $\Delta T$ for resonant odd-$\pi$ Lorentzian pulses with scaled Rabi frequencies $\Omega_0T=5,11,21,$ and $41$. The exact resonant probability is unity in every case, while the central excitation line narrows systematically as $\Omega_0T$ increases.}
\label{fig:power-profiles}
\end{figure}

The inverse-$\alpha$ law can be checked independently by direct numerical integration of the Schr\"odinger equation. Figure~\ref{fig:power-profiles} shows the narrowing central line for four resonant odd-$\pi$ pulse areas. Extracting the positive half-width from the numerical profiles for $\alpha=5,7,9,11,13,17,21,25,31,$ and $41$ gives the fit
\begin{equation}
\delta_{1/2}^{(\mathrm{num})}\simeq0.6041\,\alpha^{-1.0001}.
\label{eq:numerical-linewidth-fit}
\end{equation}
Thus the numerical exponent agrees exactly with the predicted value $-1$ within the numerical accuracy. 
The numerical half-width, $\delta_{1/2}^{(\mathrm{num})}\simeq 0.6041/\alpha$, is about $26\%$ larger than the unitarized asymptotic prediction $0.4803/\alpha$. 
The discrepancy in the coefficient reflects the fact that the unitarized DDP expression is not uniformly accurate at the half-maximum, where $\operatorname{Im}\mathcal D$ is of order unity rather than asymptotically large.

\begin{figure}[tbph]
\centering
\includegraphics[width=1\columnwidth]{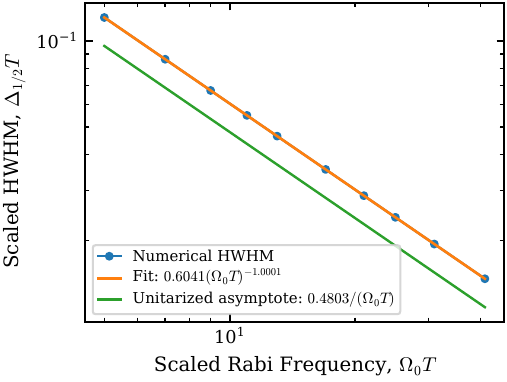}
\caption{Scaled positive half-width at half maximum $\Delta_{1/2}T$ extracted from direct numerical Schr\"odinger integration for odd-$\pi$ pulse areas and plotted against the scaled Rabi frequency $\Omega_0T$. The numerical fit, Eq.~(\ref{eq:numerical-linewidth-fit}), has exponent $-1.0001$, confirming the inverse-$\alpha$ power-narrowing law. The unitarized asymptotic prediction $0.4803/\alpha$ has the same exponent but underestimates the numerical prefactor.}
\label{fig:power-scaling}
\end{figure}

\subsection{Exact resonant limit}

At exact resonance, $\Delta=0$, the transition probability is
\be
\mathcal{P}(\Delta=0)
=\sin^2\left(\frac{A}{2}\right)
=\sin^2\left(\frac{\pi\Omega_0T}{2}\right).
\label{P-resonance}
\ee
The near-resonant unitarized DDP expression tends continuously to Eq.~(\ref{P-resonance}) as $\Delta\rightarrow0$, while the transition points themselves move to infinity. This illustrates why the exact resonant dynamics is regular even though the transition-point representation is a singular asymptotic limit.

\subsection{Comparison with the Rosen-Zener-Robiscoe conjecture\label{Sec-RZR-conjecture}}

The factorized form of the exact Rosen-Zener transition probability motivated a conjecture for more general symmetric pulse envelopes. The conjecture was examined perturbatively by Robiscoe and for short Gaussian pulses by Thomas \cite{RZ,Robiscoe1983,Thomas83}. A Robiscoe-type generalization has also been used successfully as an accurate interpolation for Gaussian, squared-hyperbolic-secant, and exponential pulses, with direct tests on an IBM quantum processor \cite{MihovVitanov2023}. In its original form, the conjecture reads
\begin{subequations}
\label{P-RZ-conjecture-general}
\begin{align}
\mathcal{P}_{\mathrm{RZc}}(\Delta)
={}&\sin^2\left(\frac{A}{2}\right)
\left|\frac{1}{A}\int_{-\infty}^{\infty}
\Omega(t)e^{i\Delta t}\,dt\right|^2,
\label{P-RZ-conjecture}\\
A={}&\int_{-\infty}^{\infty}\Omega(t)\,dt.
\label{RZ-pulse-area}
\end{align}
\end{subequations}
Thus the exact resonant pulse-area dependence is multiplied by the normalized spectral intensity of the pulse. For the Lorentzian envelope, $A=\pi\Omega_0T=\pi \alpha$ and Eq.~(\ref{Lorentzian-FT}) gives
\begin{equation}
\mathcal{P}_{\mathrm{RZc}}(\alpha,\delta)=
\sin^2\left(\frac{\pi \alpha}{2}\right)e^{-2|\delta|}.
\label{P-RZ-conjecture-Lorentzian}
\end{equation}
Equation~(\ref{P-RZ-conjecture-Lorentzian}) is exact at resonance and, for $\alpha\ll1$, reduces to the perturbative result in Eq.~(\ref{P-weak}). The Robiscoe-type extension replaces the resonant pulse area by a detuning-dependent effective area
\begin{subequations}
\label{Robiscoe-effective-area}
\begin{align}
A_{\mathrm{eff}}(\Delta)
&=\widetilde{T}\sqrt{\Omega_0^2+a\Delta^2},\\
\widetilde{T}
&=\frac{1}{\Omega_0}
\int_{-\infty}^{\infty}\Omega(t)\,dt.
\end{align}
\end{subequations}
where $a$ is a pulse-shape-dependent dimensionless parameter. For the Lorentzian pulse, $\widetilde{T}=\pi T$, and the generalized conjecture becomes
\begin{equation}
\mathcal{P}_{\mathrm{Rob}}(\alpha,\delta;a)=
\sin^2\left[\frac{\pi}{2}
\sqrt{\alpha^2+a\delta^2}\right] e^{-2|\delta|}.
\label{P-Robiscoe-Lorentzian}
\end{equation}
The original conjecture is recovered for $a=0$. At the resonant odd-$\pi$ maxima, $\alpha=2N+1$, Eq.~(\ref{P-Robiscoe-Lorentzian}) may be written as
\begin{equation}
\mathcal{P}_{\mathrm{Rob}}=
\cos^2\left\{\frac{\pi}{2}
\left[\sqrt{\alpha^2+a\delta^2}-\alpha\right]\right\} e^{-2|\delta|}.
\label{P-Robiscoe-odd-pi}
\end{equation}
For any fixed $a$, the cosine factor tends to unity at fixed $\delta$ as $\alpha\rightarrow\infty$, and the central half-width therefore approaches $|\delta_{1/2}|=(\ln2)/2$, independently of $\alpha$. Hence neither the original conjecture nor its fixed-$a$ Robiscoe generalization reproduces the inverse-$\alpha$ power narrowing derived above from the DDP action and confirmed numerically. The Lorentzian model consequently provides a clear strong-coupling limit in which the area-times-spectrum factorization, although exact at resonance and useful for several other pulse shapes, fails at the level of the linewidth scaling.

\subsection{Summary of the asymptotic regimes}

The two asymptotic regimes may be summarized as
\begin{subequations}
\label{D-summary-group}
\begin{align}
\mathcal{D}\simeq{}&i\Delta T+\frac{\pi}{4}\Omega_0T
+\frac{\pi}{16}\frac{\Omega_0^2T}{\Delta}
\notag\\
&+i\frac{\Omega_0^2T}{8\Delta}
\left[\ln\left(\frac{8\Delta}{\Omega_0}\right)-\frac12\right]
+\cdots,
\quad \beta\ll1,
\label{D-summary-small}\\
\mathcal{D}\simeq{}&\frac{\pi}{2}\Omega_0T
+(-1+i)\frac{\Gamma^2(3/4)}{\sqrt{\pi}}
 T\sqrt{\Omega_0\Delta}+\cdots,
\quad \beta\gg1.
\label{D-summary-large}
\end{align}
\end{subequations}
The weak-coupling expression yields exponentially decaying far-detuned wings and the perturbative result of Eq.~(\ref{P-weak}).
The real and imaginary parts determine the asymptotic phase and attenuation, respectively. In the near-resonant odd-$\pi$ sequence, the resulting scaling variable $\sqrt{\Omega_0|\Delta|}\,T$ implies $\Delta_{1/2}\propto\Omega_0^{-1}$, connecting the DDP analysis directly to power narrowing.

\section{Pad\'e approximations\label{Sec-Pade}}

The weak- and strong-coupling expansions derived in Secs.~\ref{sec:weak} and \ref{sec:strong} are accurate near $\beta=0$ and $\beta\rightarrow\infty$, respectively, but neither is uniform across the crossover $\beta\sim1$. We therefore construct two-point Pad\'e approximants to the dimensionless DDP action rather than to the transition probability \cite{Banks-Torres,appB-BakerGravesMorris}. Define
\begin{equation}
 x=\sqrt{\beta},\qquad
 R_\pm(x)=\frac{\mathcal{D}(\tau^\pm)}{\Delta T}.
\label{eq:Rpm-exact}
\end{equation}
It is convenient to collect the asymptotic coefficients as
\begin{subequations}
\label{eq:Rpm-asymptotic-data}
\begin{align}
s_\pm&=\pm\frac{\pi}{4},
&L_\pm&=\pm\frac{\pi}{2},\label{eq:Rpm-sL}\\
M_+&=(-1+i)C,
&M_-&=(1+i)C.\label{eq:Rpm-M}
\end{align}
\end{subequations}
The two limits to be matched can then be written compactly as
\begin{subequations}
\label{eq:Rpm-targets}
\begin{align}
R_\pm(x)&=i+s_\pm x^2
+O\!\left[x^4\ln(1/x)\right],
&&x\rightarrow0,\label{eq:Rpm-small-target}\\
R_\pm(x)&=L_\pm x^2+M_\pm x
+O(x^{-1}),
&&x\rightarrow\infty.\label{eq:Rpm-large-target}
\end{align}
\end{subequations}
The logarithmic term in Eq.~(\ref{eq:Rpm-small-target}) cannot be reproduced by a rational function of $x$ and is therefore not included in either matching scheme.

\subsection{Minimal two-point \texorpdfstring{$[3/1]$}{[3/1]} approximation\label{Sec-Pade31}}

The lowest-order rational interpolation that can satisfy the five leading conditions in Eqs.~(\ref{eq:Rpm-targets}) is a $[3/1]$ approximant. Defining
\begin{equation}
p=\frac{\pi(1+i)}{8C},
\label{eq:pade31-p}
\end{equation}
the result for the $+$ branch is
\begin{subequations}
\label{eq:pade31-explicit}
\begin{align}
R_+^{[3/1]}(x)
&=\frac{\begin{aligned}
&i-p^*x+(\pi/4)x^2\\
&\quad +(\pi/2)p x^3
\end{aligned}}{1+p x},
\label{eq:pade31-explicit-plus}\\
R_-^{[3/1]}(x)
&=-\left[R_+^{[3/1]}(x)\right]^*.
\label{eq:pade31-explicit-minus}
\end{align}
\end{subequations}
The coefficient matching is derived in Appendix~\ref{App-Pade31}. The approximation has the correct leading terms at both ends, but its first unmatched contributions are
\begin{subequations}
\label{eq:pade31-defects}
\begin{align}
R_+^{[3/1]}(x)
={}&i+\frac{\pi}{4}x^2
+\frac{\pi^2(1+i)}{32C}x^3
\notag\\
&+O(x^4),\qquad x\rightarrow0,
\label{eq:pade31-small-defect}\\
R_+^{[3/1]}(x)
={}&\frac{\pi}{2}x^2+(-1+i)Cx
\notag\\
&+i\frac{\pi-8C^2}{\pi}
+O(x^{-1}),\qquad x\rightarrow\infty.
\label{eq:pade31-large-defect}
\end{align}
\end{subequations}
Hence the minimal approximant introduces a spurious $O(x^3)=O(\beta^{3/2})$ term at weak coupling and a spurious constant at strong coupling. The latter produces an $O(1)$ error in $\operatorname{Im}\mathcal{D}$ and therefore distorts the exponentially sensitive DDP envelope. The $[3/1]$ expression is consequently useful as a compact qualitative interpolation and as a benchmark for the improved construction below, but not as a uniformly accurate approximation.

\subsection{Constrained two-point \texorpdfstring{$[4/2]$}{[4/2]} approximation\label{Sec-Pade42}}

The two defects in Eqs.~(\ref{eq:pade31-defects}) can be removed by additionally imposing the absence of the $x^3$ term at small $x$ and of the constant term at large $x$. Introduce
\begin{subequations}
\label{eq:pade42-rho-gamma}
\begin{align}
q&=8C^2-\pi,\label{eq:pade42-q}\\
\rho&=\frac{\pi C}{q}=1.0234683276\ldots,\label{eq:pade42-rho}\\
\gamma&=\frac{\pi^2}{4q}=0.9487933546\ldots.\label{eq:pade42-gamma}
\end{align}
\end{subequations}
The constrained $[4/2]$ approximation is
\begin{subequations}
\label{eq:pade42-explicit}
\begin{align}
R_+^{[4/2]}(x)&=\frac{N_+(x)}{Q_+(x)},
\label{eq:pade42-explicit-ratio}\\
N_+(x)&=i+(-1+i)\rho x+(\pi/4-\gamma)x^2
\notag\\
&\quad +(1+i)(\pi\rho/4)x^3+i(\pi\gamma/2)x^4,
\label{eq:pade42-explicit-num}\\
Q_+(x)&=1+(1+i)\rho x+i\gamma x^2,
\label{eq:pade42-explicit-den}\\
R_-^{[4/2]}(x)&=-\left[R_+^{[4/2]}(x)\right]^*.
\label{eq:pade42-explicit-minus}
\end{align}
\end{subequations}
Its derivation is given in Appendix~\ref{App-Pade42}. The denominator has no zeros on the physical interval $x\geq0$: its real part for the $+$ branch is $1+\rho x>0$, while the $-$ denominator is the complex conjugate. Direct expansion gives
\begin{subequations}
\label{eq:pade42-verified-limits}
\begin{align}
R_+^{[4/2]}(x)
={}&i+\frac{\pi}{4}x^2+O(x^4),
&&x\rightarrow0,\label{eq:pade42-small-check}\\
R_+^{[4/2]}(x)
={}&\frac{\pi}{2}x^2+(-1+i)Cx+O(x^{-1}),
&&x\rightarrow\infty.\label{eq:pade42-large-check}
\end{align}
\end{subequations}
Thus the first neglected orders agree with those of the exact asymptotic expansions. The remaining weak-coupling discrepancy begins at $O(x^4)$ because a rational function cannot reproduce the exact $x^4\ln x$ dependence.

\begin{figure}[tbph]
\includegraphics[width=0.98\columnwidth]{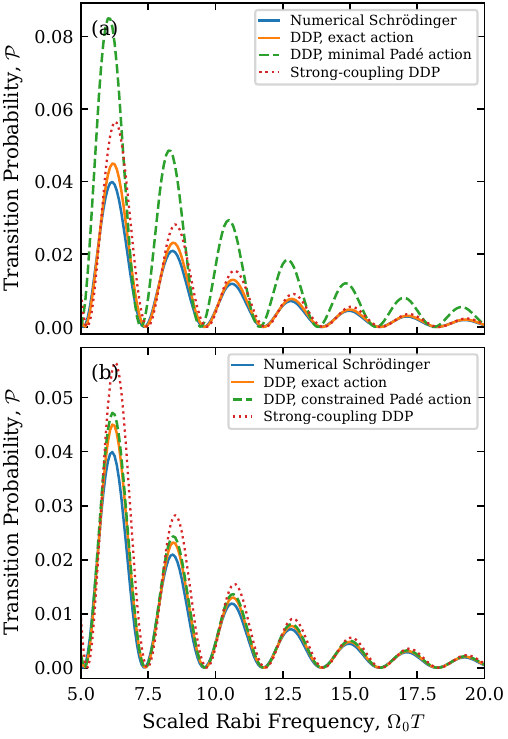}
\captionof{figure}{Transition probability versus the scaled Rabi frequency $\Omega_0T$ for the fixed scaled detuning $\Delta T=1$. Direct numerical integration of the Schr\"odinger equation is compared with the standard DDP probability evaluated using the exact contour action and Pad\'e-approximated actions. Panel (a) shows the minimal $[3/1]$ approximation, which reproduces the qualitative crossover but displays noticeable deviations in the envelope and phase. Panel (b) shows the constrained $[4/2]$ approximation, whose curve nearly coincides with the exact-action DDP result throughout the displayed crossover and strong-coupling range. The two panels use the same vertical scale, and the leading strong-coupling result is shown in both for reference.}
\label{fig:pade}
\end{figure}

For either order, let $\mathcal{D}_{[m/n]}=\Delta T R_+^{[m/n]}(\sqrt{\beta})$. The associated standard and unitarized probabilities are
\begin{subequations}
\label{eq:Ppade-group}
\begin{align}
\mathcal{P}_{[m/n]}^{(\mathrm{s})}
&=4e^{-2\operatorname{Im}\mathcal{D}_{[m/n]}}
\sin^2\!\left(\operatorname{Re}\mathcal{D}_{[m/n]}\right),
\label{eq:Ppade-standard}\\
\mathcal{P}_{[m/n]}^{(\mathrm{u})}
&=\frac{\sin^2\!\left(\operatorname{Re}\mathcal{D}_{[m/n]}\right)}
{\cosh^2\!\left(\operatorname{Im}\mathcal{D}_{[m/n]}\right)}.
\label{eq:Ppade-unitary}
\end{align}
\end{subequations}
Figure~\ref{fig:pade} compares the standard DDP probabilities obtained from the exact contour action and from both rational approximations for the representative slice $\delta=1$. The minimal $[3/1]$ expression reproduces the qualitative crossover but shows visible errors in the oscillation phase and envelope, whereas the constrained $[4/2]$ curve nearly coincides with the exact-action DDP result on this slice.

\begin{figure}[tbph]
\includegraphics[width=1\columnwidth]{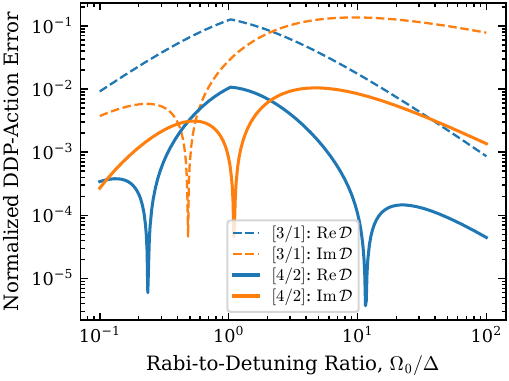}
\captionof{figure}{Normalized errors $\epsilon_X^{[m/n]}$ of the real and imaginary parts of the Pad\'e action ratio $R_+=\mathcal{D}(\tau^+)/(\Delta T)$, as defined in Eq.~(\ref{eq:pade-error-metric}), versus the Rabi-to-detuning ratio $\Omega_0/\Delta=\beta$. Dashed curves show the minimal $[3/1]$ approximation and solid curves show the constrained $[4/2]$ approximation. The additional zero-coefficient constraints reduce the maximum error over $0.1\leq\beta\leq100$ from about $13\%$ to about $1.1\%$.}
\label{fig:pade-error}
\end{figure}

To quantify the approximation of the action itself, let $X$ denote either $\operatorname{Re}$ or $\operatorname{Im}$ and define
\begin{equation}
\epsilon_X^{[m/n]}(\beta)=
\frac{\left|X\!\left[R_+^{[m/n]}(\sqrt{\beta})
-R_+(\sqrt{\beta})\right]\right|}
{\max\!\left\{1,\left|X\!\left[R_+(\sqrt{\beta})\right]\right|\right\}}.
\label{eq:pade-error-metric}
\end{equation}
Figure~\ref{fig:pade-error} plots this error of the normalized action ratio $R_+=\mathcal{D}(\tau^+)/(\Delta T)$. Over $0.1\leq\beta\leq100$, its largest real- and imaginary-part values decrease from approximately $12.7\%$ and $13.7\%$ for $[3/1]$ to approximately $1.07\%$ and $1.05\%$ for $[4/2]$. Because the full action is $\mathcal{D}=\delta R_+$ and the probability depends exponentially on its imaginary part, a small error in $R_+$ does not by itself guarantee the same relative accuracy in the transition probability for arbitrarily large $\delta$.

\section{Conclusions\label{Sec-conclusion}}

We have analyzed coherent excitation of a two-state system by a Lorentzian pulse with constant detuning. 
The associated differential equation belongs to the confluent-Heun class, although the local special-function representation does not by itself provide a simple closed-form connection formula between the physical boundary conditions at $-\infty$ and $+\infty$. 
This motivates the use of the Dykhne-Davis-Pechukas method for the post-pulse transition probability.

The Lorentzian model possesses two symmetry-related transition points in the upper complex-time plane together with poles of the coupling at $\tau=\pm i$. 
Their complex-time geometry leads to an interference form of the DDP probability in which the real part of the action controls the oscillation phase and the imaginary part controls the attenuation. 
We have taken the contour integral for the DDP action as the primary representation, avoiding branch ambiguities associated with convention-dependent elliptic-integral closed forms.

The asymptotic expansions provide a compact physical description in the two opposite regimes. 
For weak coupling or large detuning, the imaginary part approaches $\Delta T$ and the excitation wings decay as $e^{-2|\Delta|T}$; in the additional weak-field limit the DDP result reduces to the perturbative Fourier-transform expression $\mathcal{P}\simeq(\pi^2/4)(\Omega_0T)^2e^{-2|\Delta|T}$. 
For strong coupling or small detuning, the action contains the characteristic correction $T\sqrt{\Omega_0|\Delta|}$, yielding analytic expressions for the oscillation phase, envelope, and local Rabi-frequency period.

A particularly important consequence is the linewidth scaling near the odd-$\pi$ resonant maxima. 
The strong-coupling asymptotics depend on detuning through the combination $(\Omega_0T)(|\Delta|T)$, which implies $\Delta_{1/2}\propto\Omega_0^{-1}$ at fixed pulse duration. 
Within the unitarized near-resonant interpolation, the corresponding asymptotic half-width is $|\delta_{1/2}|\simeq0.480/\alpha$. Direct numerical Schr\"odinger integration gives $\delta_{1/2}\simeq0.604/\alpha$, confirming the inverse-$\alpha$ exponent while showing that the interpolation underestimates the prefactor.
The Rosen-Zener area-times-spectrum conjecture and its fixed-$a$ Robiscoe extension reproduce the exact resonant dependence and the perturbative weak-field limit, but they predict a nonzero asymptotic linewidth and therefore fail to reproduce the inverse-$\alpha$ power narrowing. The minimal $[3/1]$ Pad\'e action preserves the exact branch symmetry but generates a spurious $O(\beta^{3/2})$ weak-coupling term and a spurious strong-coupling constant. The constrained $[4/2]$ approximation removes both terms and reduces the normalized action-ratio error to approximately one percent over the tested crossover interval; the corresponding probability comparison demonstrated here is for $\delta=1$. 
The DDP analysis therefore reproduces, from a complex-time perspective, the power-narrowing law previously predicted for Lorentzian algebraic tails \cite{BoradjievVitanov2013} and later observed experimentally for Lorentzian-type pulses \cite{MihovVitanov2024}. 
This connection between transition-point interference, asymptotic action, and power narrowing is one of the main physical results of the present analysis.

\acknowledgments
This research is supported by the Bulgarian national plan for recovery and resilience, contract BG-RRP-2.004-0008-C01 (SUMMIT: Sofia University Marking Momentum for Innovation and Technological Transfer), project number 3.1.4, and by the QuantERA project QUARCKS.
GSV acknowledges support by the project QUANTNET - European Reintegration Grant (ERG) - PERG07-GA-2010-268432.

\appendix
\section{Reduction of the DDP action to elliptic integrals\label{App-elliptic}}
The DDP action on the physical sheet is
\begin{equation}
\mathcal{D}(\tau)=\Delta T\int_0^\tau
\frac{\sqrt{\beta^2+(1+\tau'^2)^2}}
{1+\tau'^2}\,d\tau'.
\label{eq:appA-start}
\end{equation}
The radicand factorizes as
\begin{equation}
\beta^2+(1+\tau'^2)^2
=(1+\tau'^2-i\beta)(1+\tau'^2+i\beta),
\label{eq:appA-factorization}
\end{equation}
so the four branch points are determined by
$1+\tau'^2=\pm i\beta$. To reduce Eq.~(\ref{eq:appA-start}) to standard elliptic integrals, define
\begin{subequations}
\label{eq:appA-substitution}
\begin{align}
n&\equiv1+i\beta,\label{eq:appA-n}\\
m&\equiv\frac{1+i\beta}{1-i\beta}
=\frac{i-\beta}{i+\beta},\label{eq:appA-m}\\
\tau'&=\sqrt{1+i\beta}\,\sinh u,\label{eq:appA-tau-sub}\\
\varphi&\equiv iu
=i\,\operatorname{arcsinh}\!\left(
\frac{\tau}{\sqrt{1+i\beta}}\right).
\label{eq:appA-phi}
\end{align}
\end{subequations}
Since $u=-i\varphi$, the factors appearing in the integrand become
\begin{subequations}
\label{eq:appA-identities}
\begin{align}
1+\tau'^2+i\beta
&=(1+i\beta)\cos^2\varphi,\label{eq:appA-id-plus}\\
1+\tau'^2-i\beta
&=(1-i\beta)(1-m\sin^2\varphi),\label{eq:appA-id-minus}\\
1+\tau'^2
&=1-n\sin^2\varphi,\label{eq:appA-id-den}\\
d\tau'&=-i\sqrt{1+i\beta}\cos\varphi\,d\varphi.
\label{eq:appA-dtau}
\end{align}
\end{subequations}
Using the parameter convention of Eq.~(\ref{elliptic-definitions}), and choosing all square roots by analytic continuation from $\tau=0$, one obtains
\begin{equation}
\begin{aligned}
\mathcal{D}(\tau)={}&
\frac{\Delta T}{\sqrt{1-i\beta}}
\Big[-(i+\beta)E(\varphi,m)+\beta F(\varphi,m)
\\[-1mm]
&\hspace{23mm}-i\beta^2\Pi(n;\varphi,m)\Big].
\end{aligned}
\label{eq:appA-general}
\end{equation}
The lower integration limit corresponds to $\varphi=0$, so no additive integration constant is required. Equation~(\ref{eq:appA-general}) is a local primitive; for the physical DDP action it must be continued along the contours specified in Sec.~\ref{Sec-transition-points}.

\subsection{Evaluation at the transition points}

For the upper-right transition point $\tau^+=\sqrt{-1+i\beta}$, define
\begin{equation}
\varphi_+
=i\,\operatorname{arcsinh}\!\left(
\sqrt{\frac{-1+i\beta}{1+i\beta}}
\right).
\label{eq:appA-phiplus}
\end{equation}
Continuation of Eq.~(\ref{eq:appA-general}) along the contour $\mathcal{C}_+$ gives
\begin{equation}
\begin{aligned}
\mathcal{D}(\tau^+)={}&
\frac{\Delta T}{\sqrt{1-i\beta}}
\Big[-(i+\beta)E(\varphi_+,m)
+\beta F(\varphi_+,m)
\\[-1mm]
&\hspace{19mm}-i\beta^2
\Pi(n;\varphi_+,m)\Big].
\end{aligned}
\label{eq:appA-Dplus}
\end{equation}
For the upper-left transition point
$\tau^-=-\sqrt{-1-i\beta}$, the same mapping gives
$\varphi_-=-\pi/2$ on the branch connected continuously to the physical contour. Because $F$, $E$, and $\Pi$ are odd functions of their amplitude,
\begin{subequations}
\label{eq:appA-complete-defs}
\begin{align}
F(-\pi/2,m)&=-K(m),\\
E(-\pi/2,m)&=-E(m),\\
\Pi(n;-\pi/2,m)&=-\Pi(n,m),
\end{align}
\end{subequations}
where $K(m)=F(\pi/2,m)$ and
$\Pi(n,m)=\Pi(n;\pi/2,m)$. Hence
\bse
\label{eq:appA-Dminus}
\begin{align}
\mathcal{D}(\tau^-)&=
\lambda\Big[(i+\beta)E(m)-\beta K(m) +i\beta^2\Pi(1+i\beta,m)\Big],
\label{eq:appA-Dminus-action}
\\
\lambda&\equiv\frac{\Delta T}{\sqrt{1-i\beta}}.
\end{align}
\label{eq:appA-lambda}
\ese
This complete-elliptic representation is algebraically simpler than Eq.~(\ref{eq:appA-Dplus}) and is the form used in Secs.~\ref{Sec-asymptotics} and \ref{Sec-Pade}. The two expressions are not independent: with the contour and branch prescription fixed above, they obey the exact symmetry
$\mathcal{D}(\tau^-)=-\mathcal{D}^*(\tau^+)$.

\subsection{Weak-coupling limit}

A direct matched expansion of the contour integral displays the origin of both the logarithm and its constant term. For the $+$ branch, introduce
\begin{equation}
w=1+\tau^2,
\qquad
R_+(\beta)\equiv\frac{\mathcal{D}(\tau^+)}{\Delta T}
=\frac12\int_1^{i\beta}
\frac{\sqrt{w^2+\beta^2}}{w\sqrt{w-1}}\,dw,
\label{eq:appA-w-form}
\end{equation}
where the contour and square roots are inherited from the physical prescription and $\sqrt{w-1}=i$ at $w=0$. Choose an intermediate scale $\beta\ll\mu\ll1$ and set $L=\mu/\beta$.

In the outer region $|w|\gg\beta$,
\begin{equation}
\frac{\sqrt{w^2+\beta^2}}{w}
=1+\frac{\beta^2}{2w^2}+O\!\left(\frac{\beta^4}{w^4}\right).
\label{eq:appA-weak-outer-integrand}
\end{equation}
Integration from $w=1$ to $w=\mu$ gives
\begin{equation}
R_{\rm out}=\sqrt{\mu-1}+\frac{\beta^2}{4}J(\mu)+\cdots,
\label{eq:appA-Rout-def}
\end{equation}
with
\begin{equation}
J(\mu)=\int_1^\mu\frac{dw}{w^2\sqrt{w-1}}
=\frac{\sqrt{\mu-1}}{\mu}
+\arctan\sqrt{\mu-1}.
\label{eq:appA-Jmu}
\end{equation}
On the physical branch, its overlap expansion is
\begin{equation}
\begin{aligned}
R_{\rm out}={}&i-\frac{i\mu}{2}-\frac{i\mu^2}{8}
+\frac{i\beta^2}{4\mu}\\
&+\frac{i\beta^2}{8}\ln\!\left(\frac{4}{\mu}\right)
-\frac{i\beta^2}{8}+\cdots .
\end{aligned}
\label{eq:appA-Rout-expansion}
\end{equation}

In the inner region set $w=\beta z$. Expanding only the regular factor $1/\sqrt{\beta z-1}$ gives
\begin{equation}
\begin{aligned}
R_{\rm in}={}&-\frac{i\beta}{2}
\int_L^i\frac{\sqrt{1+z^2}}{z}\,dz\\
&-\frac{i\beta^2}{4}
\int_L^i\sqrt{1+z^2}\,dz+\cdots .
\end{aligned}
\label{eq:appA-Rin-def}
\end{equation}
The required antiderivatives are
\begin{subequations}
\label{eq:appA-inner-antiderivatives}
\begin{align}
\int\frac{\sqrt{1+z^2}}{z}\,dz
&=\sqrt{1+z^2}
+\ln\!\left(\frac{z}{1+\sqrt{1+z^2}}\right),\\
\int\sqrt{1+z^2}\,dz
&=\frac12\left[z\sqrt{1+z^2}+\operatorname{arcsinh}z\right].
\end{align}
\end{subequations}
Using $\sqrt{1+i^2}=0$, $\ln i=i\pi/2$, and $\operatorname{arcsinh}i=i\pi/2$, one obtains
\begin{equation}
\begin{aligned}
R_{\rm in}={}&\frac{\pi}{4}\beta+\frac{i\mu}{2}
-\frac{i\beta^2}{4\mu}\\
&+\frac{\pi}{16}\beta^2+\frac{i\mu^2}{8}
+\frac{i\beta^2}{16}\\
&+\frac{i\beta^2}{8}\ln\!\left(\frac{2\mu}{\beta}\right)
+\cdots .
\end{aligned}
\label{eq:appA-Rin-expansion}
\end{equation}
All powers of the arbitrary matching scale $\mu$ cancel between Eqs.~(\ref{eq:appA-Rout-expansion}) and (\ref{eq:appA-Rin-expansion}). Their sum is
\begin{equation}
\begin{aligned}
R_+(\beta)={}&i+\frac{\pi}{4}\beta+\frac{\pi}{16}\beta^2
+i\frac{\beta^2}{8}
\left[\ln\!\left(\frac{8}{\beta}\right)-\frac12\right]
+\cdots,
\end{aligned}
\label{eq:appA-Rplus-weak-derived}
\end{equation}
which reproduces Eq.~(\ref{D-small-beta}). The second branch then follows without a separate calculation:
\begin{equation}
R_-(\beta)=-R_+^*(\beta),
\label{eq:appA-Rminus-weak-symmetry}
\end{equation}
and hence gives Eq.~(\ref{eq:Dminus-weak}). This derivation makes explicit that the logarithm originates from the overlap between the pole region $w=O(\beta)$ and the regular outer contour.

\subsection{Strong-coupling limit}

The leading strong-coupling coefficients are obtained most transparently from the contour integral itself. Put
\begin{equation}
\epsilon=\beta^{-1},\qquad
\tau=\sqrt{\beta}\,z,
\qquad z_+=\frac{\tau^+}{\sqrt{\beta}},
\label{eq:appA-strong-scaling}
\end{equation}
so that $z_+\to e^{i\pi/4}$. Equation~(\ref{D(Tc)}) becomes
\begin{equation}
R_+(\beta)=\sqrt{\beta}\int_0^{z_+}
\frac{\sqrt{1+(z^2+\epsilon)^2}}{z^2+\epsilon}\,dz.
\label{eq:appA-strong-scaled-action}
\end{equation}
Separate the singular inner contribution according to
\begin{equation}
\frac{\sqrt{1+(z^2+\epsilon)^2}}{z^2+\epsilon}
=\frac{1}{z^2+\epsilon}+G(z,\epsilon),
\label{eq:appA-strong-split}
\end{equation}
where
\begin{equation}
G(z,0)=\frac{\sqrt{1+z^4}-1}{z^2}
\label{eq:appA-G0}
\end{equation}
is regular at $z=0$. The elementary part gives
\begin{equation}
\begin{aligned}
\sqrt{\beta}\int_0^{z_+}\frac{dz}{z^2+\epsilon}
&=\beta\arctan\tau^+\\
&=\frac{\pi}{2}\beta-e^{-i\pi/4}\sqrt{\beta}
+O(\beta^{-1/2}).
\end{aligned}
\label{eq:appA-strong-singular-part}
\end{equation}
The regular remainder has the limit
\begin{equation}
\sqrt{\beta}\int_0^{z_+}G(z,\epsilon)\,dz
=I\sqrt{\beta}+O(\beta^{-1/2}),
\label{eq:appA-strong-regular-part}
\end{equation}
where
\begin{equation}
I=\int_0^{e^{i\pi/4}}
\frac{\sqrt{1+z^4}-1}{z^2}\,dz.
\label{eq:appA-I-def}
\end{equation}
Rotating the contour by $z=e^{i\pi/4}x$ gives
\begin{equation}
I=e^{-i\pi/4}\int_0^1
\frac{\sqrt{1-x^4}-1}{x^2}\,dx.
\label{eq:appA-I-rotated}
\end{equation}
An integration by parts followed by $u=x^4$ yields
\begin{equation}
\begin{aligned}
\int_0^1\frac{\sqrt{1-x^4}-1}{x^2}\,dx
&=1-2\int_0^1\frac{x^2}{\sqrt{1-x^4}}\,dx\\
&=1-\frac12 B\!\left(\frac34,\frac12\right)
=1-\sqrt{2}\,C,
\end{aligned}
\label{eq:appA-beta-integral}
\end{equation}
where
\begin{equation}
C=\frac{\Gamma^2(3/4)}{\sqrt{\pi}}
\label{eq:appA-C-derived}
\end{equation}
follows from the beta-function identity and
$\Gamma(1/4)\Gamma(3/4)=\pi\sqrt{2}$. Combining Eqs.~(\ref{eq:appA-strong-singular-part})--(\ref{eq:appA-beta-integral}) gives
\begin{equation}
R_+(\beta)=\frac{\pi}{2}\beta
+(-1+i)C\sqrt{\beta}
+O(\beta^{-1/2}),
\label{eq:appA-Rplus-strong-derived}
\end{equation}
because
$-e^{-i\pi/4}+e^{-i\pi/4}(1-\sqrt{2}C)=(-1+i)C$.
Finally,
\begin{equation}
R_-(\beta)=-\frac{\pi}{2}\beta
+(1+i)C\sqrt{\beta}
+O(\beta^{-1/2}),
\label{eq:Dminus-strong-limit}
\end{equation}
follows from the exact branch symmetry. Equations~(\ref{eq:appA-Rplus-strong-derived}) and (\ref{eq:Dminus-strong-limit}) provide an independent derivation of the strong-coupling coefficients used in Sec.~\ref{sec:strong}, including the constant $C$.


\section{Derivation of the minimal \texorpdfstring{$[3/1]$}{[3/1]} Pad\'e approximation\label{App-Pade31}}

This appendix derives Eq.~(\ref{eq:pade31-explicit}). Begin with the normalized ansatz
\begin{equation}
R_\pm^{[3/1]}(x)=
\frac{a_0^\pm+a_1^\pm x+a_2^\pm x^2+a_3^\pm x^3}
{1+b_1^\pm x}.
\label{eq:appB-pade31}
\end{equation}
Its expansions at the two ends are
\begin{subequations}
\label{eq:appB-pade31-expansions}
\begin{align}
R_\pm^{[3/1]}(x)={}&a_0^\pm
+\big(a_1^\pm-a_0^\pm b_1^\pm\big)x
\notag\\
&+\big[a_2^\pm-a_1^\pm b_1^\pm
+a_0^\pm(b_1^\pm)^2\big]x^2
+O(x^3),\label{eq:appB-pade31-small}\\
R_\pm^{[3/1]}(x)={}&\frac{a_3^\pm}{b_1^\pm}x^2
+\frac{a_2^\pm b_1^\pm-a_3^\pm}{(b_1^\pm)^2}x
+O(1).\label{eq:appB-pade31-large}
\end{align}
\end{subequations}
Matching these expressions to Eq.~(\ref{eq:Rpm-targets}) fixes the five coefficients:
\begin{subequations}
\label{eq:appB-pade31-matching}
\begin{align}
a_0^\pm&=i,
&a_1^\pm&=i b_1^\pm,\label{eq:appB-pade31-a01}\\
a_2^\pm&=s_\pm,
&a_3^\pm&=L_\pm b_1^\pm,\label{eq:appB-pade31-a23}\\
b_1^\pm&=\frac{s_\pm-L_\pm}{M_\pm}.\label{eq:appB-pade31-b1}
\end{align}
\end{subequations}
Using the coefficients in Eq.~(\ref{eq:Rpm-asymptotic-data}) gives $b_1^+=p$ and $b_1^-=p^*$, where $p$ is defined in Eq.~(\ref{eq:pade31-p}). Substitution produces Eq.~(\ref{eq:pade31-explicit}) and automatically enforces the exact branch symmetry $R_-^{[3/1]}=-[R_+^{[3/1]}]^*$. Expanding the resulting expression one order beyond the matched terms yields Eq.~(\ref{eq:pade31-defects}); no additional assumptions enter that comparison.

\section{Derivation of the constrained \texorpdfstring{$[4/2]$}{[4/2]} Pad\'e approximation\label{App-Pade42}}

The $[4/2]$ construction adds two degrees of freedom to eliminate the unmatched $x^3$ term at small $x$ and the constant term at large $x$. Write
\begin{subequations}
\label{eq:appC-PQ}
\begin{align}
P_\pm(x)&=a_0^\pm+a_1^\pm x+a_2^\pm x^2
+a_3^\pm x^3+a_4^\pm x^4,\label{eq:appC-P}\\
Q_\pm(x)&=1+b_1^\pm x+b_2^\pm x^2,\label{eq:appC-Q}\\
R_\pm^{[4/2]}(x)&=\frac{P_\pm(x)}{Q_\pm(x)}.\label{eq:appC-R}
\end{align}
\end{subequations}
At small $x$, matching Eq.~(\ref{eq:Rpm-small-target}) through $x^3$ is equivalent to
\begin{equation}
P_\pm(x)=Q_\pm(x)\big(i+s_\pm x^2\big)+O(x^4),
\label{eq:appC-small-product}
\end{equation}
which gives
\begin{subequations}
\label{eq:appC-small-matching}
\begin{align}
a_0^\pm&=i,
&a_1^\pm&=i b_1^\pm,\label{eq:appC-a01}\\
a_2^\pm&=s_\pm+i b_2^\pm,
&a_3^\pm&=s_\pm b_1^\pm.\label{eq:appC-a23}
\end{align}
\end{subequations}
At large $x$, matching Eq.~(\ref{eq:Rpm-large-target}) while eliminating the constant term requires
\begin{equation}
P_\pm(x)=Q_\pm(x)\big(L_\pm x^2+M_\pm x\big)+O(x),
\label{eq:appC-large-product}
\end{equation}
and therefore
\begin{subequations}
\label{eq:appC-large-matching}
\begin{align}
a_4^\pm&=L_\pm b_2^\pm,\label{eq:appC-a4}\\
a_3^\pm&=L_\pm b_1^\pm+M_\pm b_2^\pm,\label{eq:appC-large-a3}\\
a_2^\pm&=L_\pm+M_\pm b_1^\pm.\label{eq:appC-large-a2}
\end{align}
\end{subequations}
Define $d_\pm=s_\pm-L_\pm$. Eliminating the numerator coefficients between Eqs.~(\ref{eq:appC-small-matching}) and (\ref{eq:appC-large-matching}) gives
\begin{subequations}
\label{eq:appC-b-linear}
\begin{align}
d_\pm b_1^\pm&=M_\pm b_2^\pm,\label{eq:appC-b-linear1}\\
M_\pm b_1^\pm-i b_2^\pm&=d_\pm.\label{eq:appC-b-linear2}
\end{align}
\end{subequations}
The solution is
\begin{subequations}
\label{eq:appC-b-solution}
\begin{align}
b_1^\pm&=\frac{d_\pm M_\pm}{M_\pm^2-i d_\pm},\label{eq:appC-b1-solution}\\
b_2^\pm&=\frac{d_\pm^2}{M_\pm^2-i d_\pm}.\label{eq:appC-b2-solution}
\end{align}
\end{subequations}
The numerator coefficients follow directly from Eqs.~(\ref{eq:appC-small-matching}) and (\ref{eq:appC-a4}). For the $+$ branch, $d_+=-\pi/4$ and $M_+=(-1+i)C$, so
\begin{equation}
M_+^2-i d_+=i\left(\frac{\pi}{4}-2C^2\right).
\label{eq:appC-plus-denominator}
\end{equation}
Substitution reduces the result to Eqs.~(\ref{eq:pade42-rho-gamma}) and (\ref{eq:pade42-explicit}); the $-$ branch follows by complex conjugation. One additional expansion gives
\begin{subequations}
\label{eq:appC-next-orders}
\begin{align}
R_+^{[4/2]}(x)
={}&i+\frac{\pi}{4}x^2
+i\frac{\pi^3}{16(8C^2-\pi)}x^4
\notag\\
&+O(x^5),\qquad x\rightarrow0,
\label{eq:appC-next-small}\\
R_+^{[4/2]}(x)
={}&\frac{\pi}{2}x^2+(-1+i)Cx
\notag\\
&+O(x^{-1}),\qquad x\rightarrow\infty.
\label{eq:appC-next-large}
\end{align}
\end{subequations}
The $x^4$ coefficient in Eq.~(\ref{eq:appC-next-small}) is rational and therefore cannot reproduce the logarithmic coefficient of the exact action, but the lower-order defects of the $[3/1]$ approximation have been removed.

\bibliography{references}

\end{document}